\documentstyle[aps,prl,preprint,tighten,amssymb]{revtex}

\ifpreprintsty\renewcommand{\arraystretch}{1.2}
\else\renewcommand{\arraystretch}{1.5}\fi

\newcommand{\MAT}[1]{{\bf #1}}

\newcommand{\LAB}[1]{{\sf #1}}
\newcommand{\FUN}[1]{{\rm #1}}
\newcommand{\UL}[1]{{\underline{#1}}}
\newcommand{\D}{\FUN d}
\newcommand{\DAB}[1]{\rule[0pt]{0pt}{#1}}
\newcommand{\TFRAC}[2]{{\textstyle\frac{#1}{#2}}}
\newcommand{\HALF}{\TFRAC12}
\newcommand{\KET}[1]{|#1\rangle}
\newcommand{\BRA}[1]{\langle#1|}
\newcommand{\EMB}[1]{\mbox{\boldmath$#1$}}
\newcommand{\SEMB}[1]{\mbox{\boldmath$\scriptstyle#1$}}

\newcommand{\SO}{\EMB\sigma_\LAB{\!0}}
\newcommand{\SI}{\EMB\sigma_\LAB{\!1}}
\newcommand{\SX}{\EMB\sigma_\LAB{\!x}}
\newcommand{\SY}{\EMB\sigma_\LAB{\!y}}
\newcommand{\SZ}{\EMB\sigma_\LAB{\!z}}

\newcommand{\SSZ}{\SEMB\sigma_\LAB{\!z}}

\newcommand{\RHO}{\EMB\rho}
\ifpreprintsty\newcommand{\SUM}{{\displaystyle\sum}}%
	\else\newcommand{\SUM}{{\textstyle\sum}}\fi

\makeatletter	

\renewcommand{\section}{\@startsection{section}%
{0}{-\parindent}{\baselineskip}{0.5\baselineskip}%
{\normalfont\normalsize\bfseries}}
\makeatother

\begin{document} \draft
\enlargethispage*{12pt}

\title{%
Hadamard Products of Product Operators and
\ifpreprintsty\\\fi
the Design of Gradient-Diffusion Experiments
\ifpreprintsty\\\fi
for Simulating Decoherence by NMR Spectroscopy}
\author{\vspace{-2pt}
Timothy F. Havel${}^1$\ifpreprintsty\thanks{%
To whom correspondence should be addressed
at MIT/NED/NW14, 150 Albany St., Cambridge,
MA 02115, USA; {\tt tfhavel@mit.edu}.}\fi,
Yehuda Sharf${\,}^2$, Lorenza Viola${}^3$
and David G. Cory${}^2$
\vspace*{3pt}}
\address{
${}^1$BCMP, Harvard Medical School, 240 Longwood Ave., Boston, MA 02115 \\
${}^2$Dept. of Nuclear Engineering, MIT, 150 Albany St., Cambridge, MA 02139 \\
${}^3$Los Alamos National Laboratory, MS B256, Los Alamos, NM 87545
\vspace*{4pt}}
\date{\today} \maketitle

\vspace*{-3pt}
\begin{abstract}
An extension of the product operator formalism of NMR
is introduced, which uses the Hadamard matrix product
to describe many simple spin $1/2$ relaxation processes.
The utility of this formalism is illustrated by deriving NMR
gradient-diffusion experiments to simulate several decoherence
models of interest in quantum information processing,
along with their Lindblad and Kraus representations.
\end{abstract}

\ifpreprintsty\else
\bigskip
\begin{multicols}{2}
\fi
\section{Introduction}
The {\em product operator formalism\/} is widely
used for designing radio-frequency pulse sequences to
control the coherent evolution of multi-spin-$1/2$ systems
in liquid-state nuclear magnetic resonance (NMR) spectroscopy
\cite{VenHilbers:83,PackerWright:83,SoEiLeBoEr:83,ErnBodWok:87}.
This is a symbolic representation of the $2^N \times 2^N $ Hermitian
matrices of $N$-spin density operators, observables and Hamiltonians,
together with a collection of rules for evolving them
under radio-frequency (RF) pulses and free evolution
delays (in the Schr\"odinger, Heisenberg, or interaction
picture defined by a rotating frame \cite{ErnBodWok:87}).
This representation is obtained by expanding the matrices
versus the ``product operator'' basis consisting of all possible
$N$-fold Kronecker (or tensor) products of the usual Pauli
matrices $\SX, \SY, \SZ$ and the $2\times 2$ identity $\SI$.
Together with the simpler and more general rules
provided by the underlying geometric algebra
structure \cite{SomCorHav:98,HavelDoran:00},
the product operator formalism has also proven invaluable
in designing pulse sequences to implement a wide variety of
logic gates for quantum information processing (QIP) by NMR
\cite{CorPriHav:98,ChGeKuLe:98,JonHanMos:98,PrSoDuHaCo:99}
(see \cite{HaSoTsCo:00,CoryEtAl:00} for recent reviews).

In contrast, the general theory of NMR relaxation
\cite{ErnBodWok:87,Cowan:97} is usually developed
using the ``superoperator'' representation of linear
mappings on $N$-spin density matrices $\RHO$ relative
to the matrix basis $\KET{\UL m}\BRA{\UL m'}$ obtained
from the eigenvectors of the Zeeman Hamiltonian
$\KET{\UL m}$ ($\UL m,\UL m' \in \{0,1\}^N$).
This representation consists of the $2^{2N}\times2^{2N}$
matrices acting on the ``columnized'' density matrices versus
the corresponding vector basis $\KET{\UL m'}\KET{\UL m}$.
The Lindblad differential operator \cite{Lindblad:76}
and Kraus operator sum \cite{Kraus:83} representations,
which are becoming com\-mon\-place in quantum optics
\cite{Carmichael:99}, condensed matter physics \cite{Weiss:99},
and in foundational studies of decoherence
\cite{GiuliniEtAl:96,Percival:98,Zurek:98},
are virtually unknown to NMR spectroscopists.
The recent use of NMR gradient-diffusion methods
to \emph{simulate} theoretical decoherence processes
in experimental studies of the control of decoherence
for QIP \cite{CMPKLZHS:98,CSHSKLZ:00} has made it
desirable to develop a new representation which:
(i) blends naturally with the product operator formalism;
(ii) avoids the unwieldy use of full
$2^N\times 2^N$ superoperator matrices;
(iii) is sufficient to analyze common
NMR gradient-diffusion experiments;
(iv) describes the theoretical
decoherence models most often studied;
(v) can easily be translated into the corresponding
Lindblad, Kraus and superoperator representations.

This paper presents such a formalism,
which is based on the well-known 
\emph{Hadamard product} of matrices
\cite{Horn:90,HornJohnson:91}.
This formalism permits the representation
of relaxation processes for which the
$2^{2N}\times 2^{2N}$ superoperator matrix
is diagonal relative to some operator
({\em not\/} superoperator) basis.
As such it can handle both $T_1$ and
$T_2$ relaxation (decoherence), but not
cross-relaxation \cite{ErnBodWok:87,Cowan:97}.
In terms of Kraus operator sums,
this is equivalent to requiring
that the Kraus operators can all
be simultaneously diagonalized.
The utility of the Hadamard product formalism will
be demonstrated by giving streamlined derivations
of closed-form, time-dependent density operators
for a variety of NMR gradient-diffusion experiments.
These include collective and independent decoherence about
any axis, together with several more complex forms of
decoherence including the collective isotropic model.
When possible, the corresponding Lindblad and Kraus
operators for these experiments will also be given.
We have found, however, that the Hadamard product
permits a concise analytic description of decoherence
with arbitrary correlations among the fluctuating
fields at the different spins involved,
whereas the corresponding Lindblad and Kraus
forms entail the solution of polynomial equations,
and can be written down explicitly only in the simplest cases.

\section{Hadamard products of product operators}
Given two $M\times M'$ complex-valued matrices
$\MAT A = {[a_{mm'}]}_{m,m'=0}^{M-1,\smash{M'}-1}$ and
$\MAT B = {[\,b_{mm'}]}_{m,m'=0}^{M-1,\smash{M'}-1}$,
their {\em Hadamard product\/} is the matrix consisting
of the complex products of corresponding pairs of elements,
\ifpreprintsty
$\MAT A \odot \MAT B = {[a_{mm'}\,b_{mm'}]}_{m,m'=0}^{M-1,\smash{M'}-1}$.
\else\linebreak[4]
$\MAT A \odot \MAT B ~=~ {[a_{mm'}\,b_{mm'}]}_{m,m'=0}^{M-1,\smash{M'}-1}$.
\fi	
Like the usual matrix product, the Hadamard product
satisfies the mixed product identity with the
Kronecker product \cite{Horn:90,HornJohnson:91},
\begin{equation} \label{eq:mpf}
(\MAT A \otimes \MAT B) \odot (\MAT C \otimes \MAT D) ~=~
(\MAT A \odot \MAT C) \otimes (\MAT B \odot \MAT D) ~,
\end{equation}
but unlike the usual matrix product it is commutative.

In the following, all matrices will be expressed as linear
combinations of product operators, as described above.
Some further notations which prove useful include:
$M = 2^N$; $\MAT I = \SI \otimes\cdots\otimes \SI$
(the $M\times M$ identity); $\EMB\sigma_\mu^n = \SI \otimes\cdots\otimes
\EMB\sigma_\mu \otimes\cdots\otimes \SI$ ($\mu \in \{\LAB x,\LAB y,\LAB z\}$
with $\EMB\sigma_\mu$ in the $n$-th place, $n = 1,\ldots,N$);
${\MAT E\DAB{6pt}}_0^n = {(\MAT I + \SZ^n)/2}
= \SI \otimes\cdots\otimes (\KET0\BRA0) \otimes\cdots\otimes \SI$,
${\MAT E\DAB{6pt}}_1^n = {(\MAT I - \SZ^n)/2}
= \SI \otimes\cdots\otimes (\KET1\BRA1) \otimes\cdots\otimes \SI$
(so ${({\DAB{1.6ex}\MAT E}_0^n)}^2 = {\DAB{1.6ex}\MAT E}_0^n$,
${({\DAB{1.6ex}\MAT E}_1^n)}^2 = {\DAB{1.6ex}\MAT E}_1^n$,
and ${\DAB{1.7ex}\MAT E}_0^n{\DAB{1.7ex}\MAT E}_1^n = \DAB{2.3ex}
{\DAB{1.7ex}\MAT E}_1^n{\DAB{1.7ex}\MAT E}_0^n = \MAT O$,
the $M\times M$ zero matrix); and
\begin{equation}
\!\! \MAT E_\UL{m} \,=\, {\MAT E\DAB{6pt}}_{\delta_m^1}^1
\cdots\, {\MAT E\DAB{6pt}}_{\delta_m^N}^N
\,,\quad
\EMB\sigma_\mu^{\UL{m}} \,=\,
{(\EMB\sigma_\mu^1)}^{\delta_m^1} \cdots
{(\EMB\sigma_\mu^N)}^{\delta_m^N}
\end{equation}
where $\delta_m^n$ is the $n$-th bit in the
binary representation $\UL m$ of the integer $m$,
and the matrix powers $(\EMB\sigma_\mu^n)^0 = \MAT I$,
$(\EMB\sigma_\mu^n)^1 = \EMB\sigma_\mu^n$.
On further abbreviating $\MAT E_\UL0$ by
$\MAT E = \KET{00\cdots0}\BRA{00\cdots0}$,
a matrix $\MAT A = {[a_{mm'}]}_{m,m'=0}^{M-1,M-1}
= {[a_{mm'}]}_{m,m'=0}^{M-1}$ may be expressed using these
operations as \ifpreprintsty\else\linebreak[4]\vspace*{-6pt}\fi
\begin{equation} 
\MAT A ~=~ {\SUM_{m,m'=0}^{M-1}} ~ a_{mm'} ~
\SX^{\UL{m}}\, \MAT E\, \SX^{\UL{m}'} \,,
\end{equation}
and in particular, a diagonal matrix $\MAT D = \MAT{Diag}(\MAT d)$
($\MAT d = {[d_{mm}]}_{m=0}^{M-1}$) may be written as
\begin{equation} 
\!\!\! \MAT D ~=~ {\SUM_{m=0}^{M-1}}\, d_{mm}\, \MAT E_\UL{m}
~=~ {\SUM_{m=0}^{M-1}}\, {d\,}_{\!mm}'\, \SZ^{\UL{m}}
\end{equation}
with $\MAT d' = \MAT H\,\MAT d$, where
$\MAT H = \MAT H^1 \cdots \MAT H^N$
($\MAT H^n = (\SX^n + \SZ^n)/\sqrt2$)
is the Hadamard transform ({\em not\/} product)
of all the spins \cite{TSYKLHC:00}.

The Hadamard product of any two such product operator
expressions can be worked out from the mixed product
formula (\ref{eq:mpf}) and the multiplication table below
(in which $\EMB\sigma_0$ is the $2\times2$ zero matrix).
\begin{center}
{\bf Table 1.} \\
Hadamard multiplication table \\
for identity and Pauli matrices. \\[3pt]
\ifpreprintsty\bigskip\begin{minipage}{2.50in}
\else\begin{minipage}{1.80in}\fi
{\renewcommand{\arraystretch}{1.25}
\ifpreprintsty
\begin{tabular*}{182pt}{@{\extracolsep{10pt}}
|p{24pt}||p{24pt}|p{24pt}|p{24pt}|p{24pt}|}
\else
\begin{tabular*}{129pt}{@{\extracolsep{8pt}}
|p{15pt}||p{15pt}|p{15pt}|p{15pt}|p{15pt}|}
\fi
\hline
$~\odot$ & $\SI$ & $\SX$ & $\SY$ & $\SZ$ \\
\hline\hline
$~\SI$ & $\SI$ & $\SO$ & $\SO$ & $\SZ$ \\
\hline
$~\SX$ & $\SO$ & $\SX$ & $\SY$ & $\SO$ \\
\hline
$~\SY$ & $\SO$ & $\SY$ & $\!\!\!\!-\SX$ & $\SO$ \\
\hline
$~\SZ$ & $\SZ$ & $\SO$ & $\SO$ & $\SI$ \\
\hline
\end{tabular*}}
\end{minipage}
\ifpreprintsty\bigskip\fi
\end{center}
The essential property of Hadamard products
to be used in this Letter will now be given.
Let $\MAT A$, $\MAT B$ and $\MAT C$ be $M\times M$
complex-valued matrices with $\MAT A$, $\MAT C$ diagonal,
and let $\MAT a = \MAT{diag}(\MAT A)$, $\MAT c = \MAT{diag}(\MAT C)$
be the column vectors formed from their diagonal elements.
Then if $\MAT c^\dag$ and $\MAT C^\dag$ are the Hermitian
conjugates of $\MAT c$ and $\MAT C$, respectively:
\begin{equation} \label{eq:esp}
\MAT A\,\MAT B\,\MAT C^\dag ~=~ \left( \MAT a\, \MAT c^\dag \right)
\odot \MAT B
\end{equation}
If the matrices are sums of product operators as above,
the dyadic product $\MAT a\, \MAT c^\dag$ may
also be expressed in product operator form as
\begin{equation} \label{eq:dyad}
\MAT a\,\MAT c^\dag ~=~ \MAT A\, \MAT H\,
\MAT E\, \MAT H\, \MAT C^\dag ~.
\end{equation}

\section{Lindblad operators for gradient-diffusion}
Pulsed magnetic field gradients have many uses in NMR,
for example to estimate conditional displacement probabilities
in diverse transport phenomena \cite{Callaghan:93}.
The underlying theory is also important in
relating NMR relaxation rates to the internal
Brownian dynamics of molecules \cite{TycoEd:94}.
The experiments relevant to this Letter take
advantage of the spatial extent of the ensemble
of spin systems (molecules) in a liquid NMR sample.
A mag\-netic field gradient $\nabla B_\LAB z$ parallel
to the static field $B_\LAB z$ along the $\LAB z$-axis
causes the Zeeman precession rate of the spins to vary
linearly with their spatial $\LAB z$-coordinates,
winding the transverse ($\LAB{xy}$) magnetization
into a spiral about $\LAB z$ for which the
net transverse magnetization vanishes.
The phase coherence thus rendered unobservable can
be refocussed by either a second gradient pulse of
the opposite polarity, or by an RF $\pi$-pulse
followed by a \linebreak[2] gradient of the same polarity.
Hence to obtain true irreversible decoherence,
it is necessary to wait for diffusion to randomize the
positions of the molecules so that the correlation between
their spins' phases and $\LAB z$-coordinates is lost.
A more detailed account of this process is outside the scope
of this Letter, and may be found in Ref.~\cite{SodickCory:98}.

Let $\RHO$ be the $M\times M$ density matrix of an ensemble
of spin systems each consisting of $N$ spin $1/2$ particles,
assumed for simplicity to be noninteracting and to have the same
gyromagnetic ratio $\gamma$, which has been polarized by a static
magnetic field $B_\LAB z$ along the $\LAB z$-axis \cite{ErnBodWok:87}.
A uniform magnetic field gradient $\nabla B_\LAB z$
correlates the phases of the spins with their spatial
$\LAB z$-coordinates via the semiclassical propagator
\begin{equation} \label{eq:gee}
\!\!\! \MAT G(z) \,=\, {\SUM_{m=0}^{M-1}}\, g_{mm}(z)\,
\MAT E_\UL{m} \,=\, e^{ -\iota\, z \, k (\SSZ^1
+\cdots+ \SSZ^N) / 2 } \,.
\end{equation}
In this expression, $k = \gamma \int_0^\tau \D \tau'\;
\partial B_\LAB z \,/\, \partial z$ is the wave
number of the phase ramp along the $\LAB z$ axis,
$g_{mm}(z) = \exp( -\iota \, z \, k \, (N - 2h(m)) / 2 )$
where $h(m) = \sum_{n=1}^N\, {\DAB{1.7ex}\delta}_{\!m}^n$
is the {\em Hamming weight\/} of $m$, and $\iota^2 = -1$.
Since $\MAT G(z)\DAB{2.2ex}$ is diagonal,
its action on a density operator $\RHO$
can be written as the Hadamard product
$\MAT G(z)\,\RHO \, \MAT G^\dag(z) =
( \MAT g(z)\, \MAT g^\dag(z) ) \odot \RHO$,
and it is easily seen that
\begin{equation} 
\MAT g(z)\, \MAT g^\dag(z) ~=~
\raisebox{1pt}{\large$[$}\, e^{ \, \iota\, z\, k\, (h(m) \,-\, h(m')) }
\raisebox{1pt}{\large$]$}_{m,m'=0}^{M-1} ~.
\end{equation}

The effect of the molecular diffusion period $t$ on the
elements of $\MAT g(z)\MAT g^\dag(z)$ is to convolute
them with a Gaussian in $z$ whose variance is $D t$,
where $D$ is the diffusion constant \cite{CSHSKLZ:00,SodickCory:98}.
These may be evaluated via Fourier transform methods to
yield the corresponding {\em phase damping matrix}
\begin{equation} \label{eq:dph}
\MAT D(t) ~=~ \raisebox{1pt}{\Large$[$}\,
e^{ -\left( k\hspace{0.5pt} (h(m) - h(m')) \right)^2 Dt\, }
\raisebox{1pt}{\Large$]$}_{m,m'=0}^{M-1} ~.
\end{equation}
Following a refocusing gradient pulse
of equal magnitude and opposite polarity,
the loss of coherence due to diffusion is now given
(assuming no coherent evolution) by the time-dependent density
matrix $\RHO(t) = \MAT D(t) \odot \RHO$ (so $\RHO(0) = \RHO$).
Differentiation yields $\dot{\RHO}(t) = -\MAT R \odot \RHO(t)$,
where $\MAT R = {[\,r_{mm'}]}_{m,m'=0}^{M-1}$ is the rate matrix
\pagebreak[2]
\begin{equation} \label{eq:pdr}
\MAT R ~=~ \raisebox{0pt}{\large$[$} \! \left(
k\hspace{1pt} (h(m) - h(m')) \DAB{9pt} \right)^2 \!D
\hspace{1pt}\raisebox{0pt}{\large$]$}_{m,m'=0}^{M-1} ~.
\end{equation}
Conversely, integration yields $\MAT D(t) = \exp_\odot(
-\MAT R\, t ) = {[ \exp( -r_{mm'\,}t ) ]}_{m,m'=1\,}^{M-1}$.
As $t \rightarrow \infty$ all elements of $\MAT D(t)$
and hence $\RHO(t)$ vanish save those with $h(m) = h(m')$.
Those with $m \ne m'$ represent coherences between states
with equal angular momentum about the $\LAB z$-axis, which are
called {\em zero-quantum coherences\/} in NMR \cite{ErnBodWok:87}.

By expanding the square in Eq.~(\ref{eq:pdr}),
it is easily seen that $\MAT R$ can be written as
\begin{equation} 
\MAT R ~=~ \HALF\left( \DAB{9pt} \smash{
\MAT 1\, {(\EMB \ell \odot \EMB \ell)}^{\sf T} \,+\,
(\EMB \ell \odot \EMB \ell)\, \MAT 1^{\,\sf T} }
\right) \,-\, \EMB \ell\, \EMB \ell^{\,{\sf T}} ~,
\end{equation}
where $\MAT 1$ is a length $M$ vector of $1$'s,
$\EMB \ell = {[\,k\,  \sqrt{2D}\, h(m) ]}_{m=0}^{M-1}$,
and $\MAT 1^\top$, $\EMB{\ell}^\top$ are their transposes.
The application of Eq.~(\ref{eq:esp}) thus yields
\begin{equation} \label{eq:gdl}
\dot{\RHO}(t) ~=~ \MAT L \, \RHO(t) \, \MAT L -
\HALF\,\MAT L^{\,2} \,\RHO(t) - \HALF\,\RHO(t)\,\MAT L^{\,2} ~,
\end{equation}
which is the desired Lindblad master equation
with a single real diagonal Lindblad operator
$\MAT L = \MAT{Diag}(\,\EMB \ell\,) = \MAT L^{\dag}$
\cite{Lindblad:76,Percival:98,PresKitaev:98}.
Because $\MAT{diag}(\SZ^1 +\cdots+ \SZ^N)
= [ N - \DAB{2.2ex}
\ifpreprintsty\else\linebreak[3]\fi
{2h(m) ]}_{m=0}^{M-1} \DAB{2.3ex}$,
this can also be expressed as $\MAT L = k \sqrt{D / 2}
\ifpreprintsty\,\else\linebreak[3]\fi
(N \, \MAT I - \SZ^1 -\cdots- \SZ^N)\DAB{2.2ex}$,
and because the action of $\MAT L$ on $\RHO(t)$
is unchanged by its overall sign or by adding
on a multiple of the identity $\MAT I$,
it can be further simplified to\linebreak[2]
\begin{equation} \label{eq:colz}
\MAT L ~=~ k\, \sqrt{D / 2}\, (\SZ^1 +\cdots+ \SZ^N) ~.
\end{equation}

This result is readily generalized to cases
in which the transverse magnetization from each
type of spin has its own wave number $k^n$.
Then the propagator in (\ref{eq:gee}) becomes
$\MAT G(z) = \exp( -\iota z (k^1 \EMB\sigma%
_\LAB z^1 +\cdots+ k^N \EMB\sigma_\LAB z^N))$,
and an essentially identical derivation
leads to Eq.~(\ref{eq:gdl}) with
\begin{equation} 
\MAT L ~=~ \sqrt{D / 2} \, (k^1\,\!\SZ^1 \,+\,\cdots\,+\, k^N\,\!\SZ^N) ~.
\end{equation}
Decoherence processes of this kind occur naturally
in heteronuclear gradient-diffusion experiments,
and can be obtained in homonuclear by a sequence
of gradient pulses interspersed with $\pi$-pulses.
If these refocus certain spins so that $k^n = 0$,
the sequence applies the decoherence
{\em selectively\/} to the remaining spins.
This analysis can be further generalized to the
{\em conditional gradient operations\/} introduced
in \cite{ShaHavCor:00a}, where the gradient pulses
are interspersed with more complex RF pulse sequences
and delays which implement conditional quantum logic
gates such as the controlled-NOT \cite{SomCorHav:98}.

To illustrate such ``conditional decoherence'',
consider a two-spin system subjected to a pair
of gradient pulses selective for the first spin and
interspersed by controlled-NOT's $\MAT S^{1|2} = \SX^1
\MAT E_1^2 + \MAT E_0^2$ to spin $1$ conditional on $2$,
giving a net propagator \cite{ShaHavCor:00a} \pagebreak[2]
\begin{equation} \begin{array}{rcl} \label{eq:cg1}
\MAT G(z) &=&
e^{ -\iota\, z \, k_2\, \SSZ^1 / 2 } \,\MAT S^{1|2}\,
e^{ -\iota\, z \, k_1\, \SSZ^1 / 2 } \,\MAT S^{1|2}
\ifpreprintsty\vspace{6pt}\fi\\
&=& e^{ -\iota\, z\, \SSZ^1
(k_1 \,+\, k_2\, \SSZ^2) / 2} ~,
\end{array} \end{equation}
where the subscripts on the $k$'s now specify the
temporal order of the corresponding gradient pulses.
A similar derivation then gives the Lindblad operator
\begin{equation} 
\MAT L ~=~ \sqrt{D / 2}\; \SZ^1 (k_1 \MAT I + k_2\, \SZ^2)
\end{equation}
which for $k_1 = k_2$ selectively decohers
all off-diagonal elements of $\RHO$ except
$\SX^1 \MAT E_1^2 \leftrightarrow \rho_{13} = \rho_{31}^*$.
As another example, take a three-spin system and
substitute the controlled-NOT's in Eq.~(\ref{eq:cg1})
by Toffoli gates $\MAT T^{1|23} = \MAT I - \MAT
E_1^2 \MAT E_1^3 + \SX^1 \MAT E_1^2 \MAT E_1^3$.
Then the Lindblad operator
\begin{equation} 
\MAT L \,=\, \sqrt{D / 2}\, \SZ^1\! \left( k_1 \MAT I
+ k_2 (\MAT I + \SZ^2 + \SZ^3 - \SZ^2\SZ^3) \right)
\end{equation}
decohers the off-diagonal elements
$\SX^1 \MAT E_1^2 \MAT E_1^3 \leftrightarrow
\rho_{37} = \rho_{73}^*$ only if $k_1 \ne k_2$,
and otherwise creates this pseudo-pure
state directly from $\RHO = \SX^1$.
In general, the Lindblad operator will be
proportional to the effective Hamiltonian of the
gradient propagator preceding the diffusion period.

Thus far the discussion has been restricted to a {\em single\/}
diffusion period, so that the decoherence, although possibly
selective or conditional, acts collectively on all the affected spins.
The use of multiple diffusion periods permits implementation of
arbitrary correlations including the independent case \cite{CSHSKLZ:00}.
The phase damping matrix in this latter case is the sum of those
for each spin individually, leading to the Lindblad master equation
\begin{equation} \label{eq:ime}
\!\!\! \dot{\RHO}(t) ~=~
\ifpreprintsty \SUM_{n=1}^N \else \text{\small$\sum_{n=1}^N$} \fi
\left( \MAT L_n\, \RHO(t) \, \MAT L_n - \HALF {\MAT L_n}^2\,
\RHO(t) - \HALF \RHO(t)\, {\MAT L_n}^2 \right)
\end{equation}
with $\MAT L_n = \sqrt{D / 2}\,\, k^n\, \SZ^n$ for $n = 1,\ldots,N$.
A general formula for arbitrary correlations is likewise expected
to involve $N$ Lindblads, but appears quite complicated.
The result for a two-spin system, however, can be given as
\begin{equation} 
\MAT L_1 ~=~ a_+ \SZ^1 + b_- \SZ^2 ~,\quad
\MAT L_2 ~=~ a_- \SZ^1 + b_+ \SZ^2 ~,
\end{equation}
with
${a_\pm}^2 = \HALF R^1 \pm S\, \Delta$,
${b_\pm}^2 = \HALF R^2 \pm S\, \Delta$,
and
\begin{equation} 
\Delta ~=~ \sqrt{ \frac{R^1R^2 \,-\, {(S)}^{\smash2} }
{4{(S)}^2 + {(R^1 - R^2)}^2}} ~.
\end{equation}
Here, $R^n = {(k^n)}^2 D / 2$ are the
selective decoherence rates for the two spins,
and $S$ with $-\sqrt{R^1R^2} \le S \le \sqrt{R^1R^2}$
is a measure of the correlation in their mutual decoherence.
These equations are indeterminate if $R^1 - R^2 = 0 = S$,
and care must be taken in the choice of signs
for the square roots of ${a_\pm}^2$ and ${b_\pm}^2$.
Specifically, if $R^1 \ge R^2$ and $S^2 > R^2(R^1 - R^2)/2$
all four roots are positive, whereas $b_-$ is negative if
$S^2 < R^2(R^1 - R^2)/2$, and if $R^1 \le R^2$ then all
four roots are positive unless $S^2 < R^1(R^2 - R^1)/2$,
in which case the negative square root is taken for $a_{-\,}$;
if $S = 0$ then Eq.~(\ref{eq:ime}) is used instead.

In closing this section, we note that standard results
in the theory of Kronecker products \cite{HornJohnson:91}
show that the $M^2\times M^2$ matrix $\EMB{\mathcal L}$
of a Lindbladian superoperator of the form given in
Eq.~(\ref{eq:ime}) (with any Hermitian $\MAT L_n$) is
\begin{equation}
\EMB{\mathcal L} ~=~ \HALF\, \SUM_{n=0}^N \left( \MAT
L_n^\top \otimes \MAT I - \MAT I \otimes \MAT L_n \right)^2 ~,
\end{equation}
and that this is diagonal whenever the $\MAT L_n$ are.

\section{Kraus operator sums and correlated
\ifpreprintsty\else\hfil\\\fi decoherence}
Performing an eigenvalue decomposition of the phase
damping matrix (\ref{eq:dph}) gives the integrated
evolution equation in the standard Kraus operator sum form,
\begin{equation}  \begin{array}{rcl}
\!\!\!\! \RHO(t) \,=\, \RHO \odot \MAT D(t)
&=& \RHO \odot {\SUM_{m=0}^{M-1}}\,
\MAT k_m(t)\, \kappa_m(t)\, \MAT k_m^{\sf T}(t) \!
\ifpreprintsty\vspace{12pt}\fi\\[3pt]
&=& {\SUM_{m=0}^{M-1}}\,
\MAT K_m(t)\, \RHO\, \MAT K_m(t) ~,
\end{array} \end{equation}
where the real diagonal matrices $\MAT K_m(t) \!=\!
\sqrt{\kappa_m(t)}$ \linebreak[1] $\MAT{Diag}(\MAT k_m(t))$
are in general complicated functions of time $t$.
The only assumption made here is that $\MAT D(t)$
is pos\-itive-semidefinite for all $t$, as it must
if $\RHO(t)$ is to be positive-semidefinite
for all $t$ and inital states $\RHO = \RHO(0)$.

Due to its algebraic complexity, in general
the Kraus form can be obtained only by numerically
diagonalizing $\MAT D(t)$ at each time point. 
For independent decoherence at the different spins, however,
the rates $r_{mm'}$ are easily seen to be proportional to the
squared Hamming distances $\sum_{n=1}^N (\delta_m^n -
\delta_{m'}^n)^2$, rather than to $(h(m) - h(m'))^2$.
This leads to the Kraus form \cite{PresKitaev:98}
\begin{equation} 
\RHO(t) ~=~ 2^{-N}\, {\SUM_{m=0}^{M-1}}\, \kappa_m(t)\,
\SZ^{\UL{m}}\,\! \RHO\, \SZ^{\UL{m}} ~,
\end{equation}
where $\kappa_m(t) = \prod_{n=1}^N (1 + (-1)^{\delta_m^n} p^n(t))$
for the one-spin survival probabilities $p^n(t) = \exp( -tR^n )$.
The case of collective decoherence appears substantially more difficult,
and the eigenvalues $\kappa_m(t)$ involve radicals even for just two spins.
Assuming as in Eq.~(\ref{eq:dph}) that the one-spin survival
probabilities are both equal to $p = p(t)$, a simple algebraic form,
not based on diagonalization, exists and is given by
\begin{equation}  \begin{array}{rcl} \!
\RHO(t) &=& \TFRAC14 \left(\! (1 \!+\! p) \MAT I - p\, \SZ^1\SZ^2 \right)
\!\RHO\! \left( (1 \!+\! p) \MAT I - p\, \SZ^1\SZ^2 \right) \\
&& +\, \TFRAC14 (1 - {(p)}^2) \left( \MAT I + \SZ^1\SZ^2 \right)
\RHO \left( \MAT I + \SZ^1\SZ^2 \right) \\
&& +\, \HALF (1 - {(p)}^4) \left( \SZ^1 + \SZ^2 \right)
\RHO \left( \SZ^1 + \SZ^2 \right)
\end{array} \end{equation}
This does not seem to extend to larger numbers of spins.

The \emph{extended} Kraus (diagonal) operator sum form,
\begin{equation} 
\RHO(t) ~=~ {\SUM_{m,m'=0}^{M-1}}\; c_{mm'}(t)\,
\SZ^{\UL{m}} \RHO\; \SZ^{\UL{m}'} ~,
\end{equation}
turns out to be algebraically simpler, and
(given that $[c_{mm'}(t)]$ is positive-semidefinite)
is readily converted into the above
standard form by diagonalization.
To derive the extended form with
arbitrary correlations between two spins,
Eq.~(\ref{eq:dyad}) is used to express the microscopic
effect of an arbitrary decoherence process as
\begin{equation}  \begin{array}{r}
\!\!\!\!\! \left(
e^{-\iota\,z(k^1\SSZ^1+k^2\SSZ^2)/2}
\,\MAT H\,\MAT E\, \MAT H\,
e^{\,\iota\,z(k^1\SSZ^1+k^2\SSZ^2)/2}
\right) \odot \RHO \; \\
=~ \TFRAC14 \left( \MAT I
+ e^{-\iota\,z\,k^1\SSZ^1}\, \SX^1
+ e^{-\iota\,z\,k^2\SSZ^2}\, \SX^2
\ifpreprintsty+\else
\right. \\ \left. \qquad\qquad +\,
\fi
e^{-\iota\,z\,(k^1\SSZ^1
+ k^2\SSZ^2)}\, \SX^1\SX^2
\right) \odot \RHO \,,
\end{array} \end{equation}
where the exponentials on the right-hand
side have been commuted past each term of
$\MAT H\,\MAT E\,\MAT H = (\MAT I + \SX^1 + \SX^2 + \SX^1\SX^2)/4$,
changing the sign of their arguments as appropriate.
On combining the pulse sequence which implements
this propagator with one or more diffusion periods,
the resulting macroscopic average is given by \cite{CSHSKLZ:00}
\pagebreak[2]
\begin{equation} \label{eq:ishad}
\ifpreprintsty
\RHO(t) ~=~
\TFRAC14\left( \MAT I + e^{-R^1t}\, \SX^1 + e^{-R^2t}\, \SX^2 +
e^{-(R^1+R^2+S\SSZ^1\SSZ^2)t}\, \SX^1\SX^2 \right) \odot \RHO ~,
\else	
\begin{array}{rcl}
\RHO(t) &=&
\TFRAC14\left( \MAT I + e^{-R^1t}\, \SX^1 +
e^{-R^2t}\, \SX^2
\right. \\ && \left. \quad +\,
e^{-(R^1+R^2+S\SSZ^1\SSZ^2)t}\,
\SX^1\SX^2 \right) \odot \RHO ~,
\end{array}
\fi 
\end{equation}
where the rates $R^1$, $R^2$ and $S$ are as given previously.

The Hadamard multiplication table ({\bf Table 1}) shows the
identity term in (\ref{eq:ishad}) annihilates all but the
$\MAT I$, $\SZ^1$, $\SZ^2$ and $\SZ^1\SZ^2$ components of $\RHO$,
while the $\SX^1$ term annihilates all but the $\SX^1$,
$\SY^1$, $\SX^1\SZ^2$ and $\SY^1\SZ^2$ components, etc.
These components are isolated by the projections parallel
($+$) and perpendicular ($-$) to the $\LAB z$-axis,
\ifpreprintsty
\begin{equation} \label{eq:prj}
\RHO_{\epsilon_1\epsilon_2} ~=~
\else $\RHO_{\epsilon_1\epsilon_2} =$
\begin{equation} \label{eq:prj} \fi
\TFRAC14 (\RHO + \epsilon_1\, \SZ^1\RHO\,\SZ^1 + \epsilon_2\,
\SZ^2\RHO\,\SZ^2 + \epsilon_1\epsilon_2\, \SZ^1\SZ^2\RHO\,\SZ^1\SZ^2)
\end{equation}
($\epsilon_1,\epsilon_2\in\{\pm1\}$),
which enables the evolution equation to be
written without using the Hadamard product as
\begin{equation} \label{eq:nohad}
\ifpreprintsty 
\RHO(t) ~=~ \RHO_{++} + e^{-R^1t}\,
\RHO_{-+} + e^{-R^2t}\, \RHO_{+-}
\,+\, e^{-R^1t}\, e^{-R^2t}\,
e^{-S\SSZ^1\SSZ^2t}\, \RHO_{--} ~.
\else 
\begin{array}{rcl}
\RHO(t) &=& \RHO_{++} + e^{-R^1t}\,
\RHO_{-+} + e^{-R^2t}\, \RHO_{+-} \\
&& +\, e^{-R^1t}\, e^{-R^2t}\,
e^{-S\SSZ^1\SSZ^2t}\, \RHO_{--} ~.
\end{array}
\fi 
\end{equation}
Remarkably, Eqs.~(\ref{eq:ishad}--\ref{eq:nohad})
extend easily to \emph{any} number of spins.
The last term in (\ref{eq:ishad}), previously derived in
\cite{CSHSKLZ:00}, can be cast in a more symmetric form by
expanding $\exp(-S\SZ^1\SZ^2t) = \cosh(St) - \SZ^1\SZ^2\sinh(St)$
and applying it to $\RHO_{--}$ in Eq.~(\ref{eq:prj}),
leading eventually to the two-spin extended Kraus operator sum
\ifpreprintsty
\begin{equation} \begin{array}{r}
8 \RHO(t) ~=~
\else 
$8 \RHO(t) =$
\begin{equation} \begin{array}{r}
\fi 
a\, (\SZ^1\RHO\,\SZ^2 + \SZ^2\RHO\,\SZ^1 -
\SZ^1\SZ^2\RHO - \RHO\,\SZ^1\SZ^2) + b_{++}^{+} \, \RHO \\
+\; b_{-+}^{-} \, \SZ^1\RHO\,\SZ^1 + b_{+-}^{-} \, \SZ^2\RHO\,\SZ^2
+ b_{--}^{+} \, \SZ^1\SZ^2\RHO\,\SZ^1\SZ^2
\end{array} \end{equation}
where $a = p^1p^2(1\,-\,(q)^2)/q$ and $b_{\epsilon_1\epsilon_2}^{\epsilon_3}
= 2(1+\epsilon_1 p^1)(1+\epsilon_2 p^2) + \epsilon_3 p^1p^2(q-1)^2/q$
($\epsilon_1,\epsilon_2,\epsilon_3 \in \{\pm1\}$),
with $p^1 = \exp(-R^1t)$, $p^2 = \exp(-R^2t)$, $q = \exp(-St)$.

Finally, by sandwiching a gradient between
RF pulses for a rotation and its inverse,
the spins can be decohered about any axis.
For example, a Hadamard transform may be used to
obtain collective decoherence about the $\LAB x$-axis as
\begin{equation} 
\RHO(t) ~=~ \MAT H \left( \MAT D(t) \odot
(\MAT H\,\RHO\,\MAT H) \right) \MAT H ~,
\end{equation}
where $\MAT D(t)$ is as given in Eq.~(\ref{eq:dph}).
Isotropic collective decoherence about all three axes
is obtained by applying three identical gradients,
two of which are sandwiched between RF pulses for
$(\pi/2)$-rotations about $\LAB x$ and $\LAB y$,
and each followed by equal diffusion periods.
The Lindblad operators in this case are that in
Eq.~(\ref{eq:colz}) together with the two obtained
by replacing $\SZ$ by $\SX$ and $\SY$ throughout.
The integrated form may be written compactly as
\begin{equation}
\ifpreprintsty
\RHO(t) ~=~ \MAT{ZHZ}^\dag \text{\Large$($} \MAT D(t) \odot
\text{\Large$($} \MAT{ZHZ}^\dag \MAT H\, \text{\large$($} \MAT D(t)
\odot \text{\large$($} \MAT H\, ( \MAT D(t) \odot \RHO )
\MAT H\text{\large$)$} \MAT{HZHZ}^\dag\, \text{\large$)$}
\text{\Large$)$} \text{\Large$)$} \,\MAT{ZHZ}^\dag ~,
\else \begin{array}{rl}	
\RHO(t) ~= & \MAT{ZHZ}^\dag \text{\Large$($} \MAT D(t) \odot
\text{\Large$($} \MAT{ZHZ}^\dag \MAT H\, \text{\large$($} \MAT D(t)
\;\odot \\ &\;
\text{\large$($} \MAT H\, ( \MAT D(t) \odot \RHO )
\MAT H\text{\large$)$} \MAT{HZHZ}^\dag\, \text{\large$)$}
\text{\Large$)$} \text{\Large$)$} \,\MAT{ZHZ}^\dag ~,
\end{array} \fi
\end{equation}
where $\MAT Z = \exp( -\iota\pi(\SZ^1+\cdots+\SZ^N)/4) )$.

\section{Conclusions}
It has been shown that NMR gradient-diffusion methods
enable precise implementations of the adiabatic decoherence
processes most often studied in QIP \cite{PresKitaev:98},
and that the Hadamard product formalism is a simple
and efficient means of analyzing such processes
(regardless of their underlying physical mechanism).
Nonadiabatic relaxation of $\RHO$ towards the
equilibrium density matrix $\RHO_\LAB{eq}$ can
also be described using the Hadamard product,
simply by decohering the diagonal part of
$\RHO - \RHO_\LAB{eq}$ about the $\LAB x$-axis.
Using the fact that $\MAT I \odot \MAT X =
\MAT{Diag}(\MAT{diag}(\MAT X))$ for any $\MAT X$,
this leads to
\begin{equation}
\RHO(t) ~=~ \MAT H\! \left( \MAT D(t) \odot \left( \MAT H\! \left( \MAT I
\odot (\RHO - \RHO_\LAB{eq}) \right) \!\MAT H \right) \right) \!\MAT H +
\RHO_\LAB{eq} ~,
\end{equation}
where $\MAT D(t) = \prod_{n=1}^N \exp_\odot(-t (\MAT I + \SX^n) / T_1^n)$
for independent decoherence (although any of
the forgoing decoherence models could be used).
Nonadiabatic $T_2$ processes can similarly be described
using complex-valued phase damping matrices $\MAT D(t)$.

The ability to simulate complicated decoherence
processes by gradient-diffusion, together
with modulation of the natural decoherence
processes operative in NMR \cite{TSSKLHC:00},
should sig\-nificantly enhance the utility of NMR
as a testbed for the development of more powerful
quantum information processors \cite{CoryEtAl:00}.
As a means of designing such experiments,
the Hadamard formalism is effectively limited to
operations for which the propagator $\MAT G(z)$
can be diagonalized by a unitary matrix that is
independent of $z$, since otherwise the algebra and
integrations will not usually be analytically tractable.
Some decoherence processes, e.g.~those involving
cross-relaxation \cite{ErnBodWok:87,Cowan:97},
cannot be efficiently described by Hadamard products,
and it is not known if they can be obtained by gradient-diffusion.
Further work on the universality \cite{LloydViola:00,BCCKLZ:00},
\emph{and} complexity, of these simulations is needed.

\ifpreprintsty\bigskip\fi
\section*{Acknowledgements}
This work was funded by DARPA/MTO
through ARO grant DAAG55-97-1-0342;
L.V.~was supported by funds
from the NSA and DOE.

\ifpreprintsty\else\vspace{-\baselineskip}\fi
\bibliographystyle{unsrt}
\bibliography{../../self,../../nmr,../../phys,../../math}

\ifpreprintsty\else\end{multicols}\fi
\end{document}